# Global 21-cm Cosmology from the Farside of the Moon

*Authors:* Jack O. Burns (University of Colorado Boulder, CU-Boulder), Stuart Bale (UC-Berkeley), Richard Bradley (NRAO), Z. Ahmed (DOE/SLAC), S.W. Allen (Stanford/SLAC), J. Bowman (ASU), S. Furlanetto (UCLA), R. MacDowall (NASA GSFC), J. Mirocha (McGill University), B. Nhan (NRAO), M. Pivovaroff (DOE/SLAC), M. Pulupa (UC Berkeley), D. Rapetti (NASA ARC-USRA/CU-Boulder), A. Slosar (Brookhaven National Laboratory), K. Tauscher (CU-Boulder)

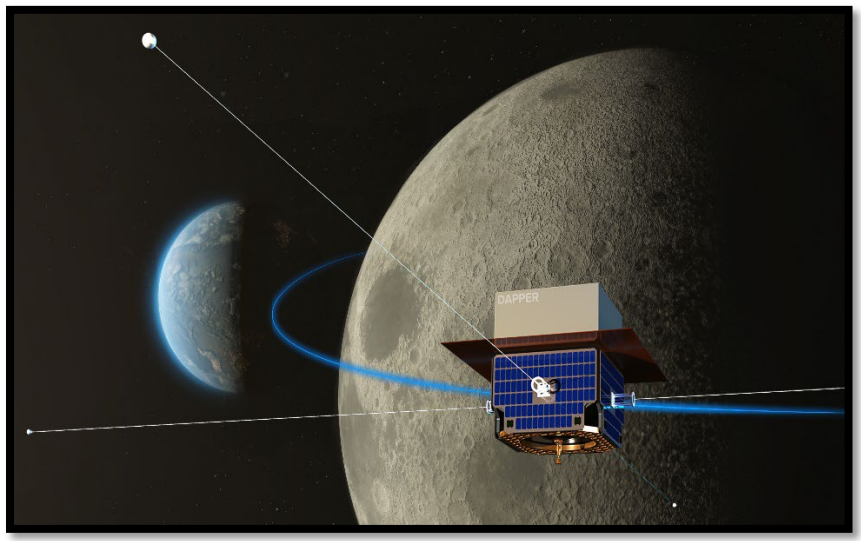

Artist concept of a Global 21-cm Cosmology telescope on a small satellite (DAPPER = Dark Ages Polarimetry PathfindER) in low lunar orbit. The telescope operates between ~10-110 MHz and takes data only above the radio-quiet lunar farside. The telescope will explore the unobserved Dark Ages and Cosmic Dawn of the early Universe using the hyperfine transition of neutral hydrogen redshifted into the VHF radio band.

## ABSTRACT – Focus Area 1: lunar farside radio telescope to explore the early universe

We describe how to open one of the last unexplored windows to the cosmos, the Dark Ages and Cosmic Dawn, using a simple low radio frequency telescope from the stable, quiet lunar farside to measure the Global 21-cm spectrum. The frontier in cosmology is the period between the onset of density fluctuations in baryons (observed via the Cosmic Microwave Background, CMB) and when the first stars, galaxies, and black holes form. This epoch marks the birth of structural complexity in the Universe, yet it remains an enormous unobserved gap in our knowledge of the Universe. Standard models of physics and cosmology are untested during this critical epoch. The messenger of information about this period is the 1420 MHz (21-cm) radiation from the hyperfine transition of neutral hydrogen, which fills the intergalactic medium, Doppler-shifted to low radio astronomy frequencies by the expansion of the Universe. The Global or all-sky averaged 21-cm spectrum uniquely probes the cosmological model during the Dark Ages plus the evolving astrophysics during Cosmic Dawn, yielding constraints on the characteristics of the first stars, on accreting black holes, and on more exotic physics such as dark matter-baryon nongravitational interactions. These observations address the question: **Will the measured neutral hydrogen spectrum redefine the standard cosmological model and reveal new physics?** This matches well with both NASA and DOE science drivers to "understand how the Universe works at a fundamental level". A single low frequency radio telescope, in either lunar orbit or on the farside surface, can measure the Global spectrum between ~10-110 MHz because of the ubiquity of neutral hydrogen during this epoch. Precise characterizations of the telescope and its surroundings are required to detect this weak, isotropic emission of hydrogen amidst the bright "foreground" Galactic radiation. We describe how two antennas will permit observations over the full frequency band: a pair of orthogonal wire antennas and a 0.3-m$^3$ patch antenna. A four-channel correlation spectropolarimeter, commonly used in radio astronomy, forms the core of the detector electronics for the telescope. Technology challenges that are well-suited to DOE/NASA collaborations include advanced calibration techniques to disentangle covariances between a bright foreground and a weak 21-cm signal, using techniques similar to those applied to the CMB, thermal management for temperature swings of >250℃, and efficient power to allow operations through a two-week lunar night. This simple telescope sets the stage for an interferometric array on the lunar farside to measure the Dark Ages power spectrum.

# 1. Low Radio Frequency Observations of the Early Universe

In this RFI response, the scientific impact of cosmological observations from the Moon is described. By measuring the Global or all-sky averaged radio frequency spectrum from ∼10-110 MHz arising from redshifted neutral hydrogen, we can explore one of the last frontiers of the Universe—the *Dark Ages* and *Cosmic Dawn*. The lunar farside is distinctively stable and radio-quiet enabling precision spectral measurements to (1) test the standard models of physics and cosmology in a new epoch, (2) probe for potential new and exotic physics (e.g., dark matter decay, early Dark Energy), and (3) place constraints on the properties of the first generation of stars and galaxies to ignite in the early Universe. These topics are well-aligned with DOE HEP and NASA Astrophysics science drivers.

## 1.1 KEY SCIENCE OPPORTUNITY

There is an enormous gap in our understanding of the early Universe between the epochs of Recombination and Reionization. Figure 1 illustrates this knowledge gap. After the Big Bang, the Universe was hot, dense, and nearly homogeneous. As the Universe expanded, the material cooled, condensing after ~400,000 years (redshift z~1100) into neutral atoms (Recombination), freeing the Cosmic Microwave Background (CMB) photons. There were no compact sources of radiation during this pre-stellar "Dark Ages". The baryonic content consisted primarily of neutral hydrogen (HI) distributed throughout the intergalactic medium. Some ∼50-100 million years later, gravity initiated the formation of the first luminous objects – stars, black holes, and galaxies – which ended the Dark Ages and commenced the "Cosmic Dawn" [1]. These mostly metal-free first stars (Population III or Pop III) likely differed dramatically from stars we see nearby (Pop I and II), as they formed in vastly different environments [2].

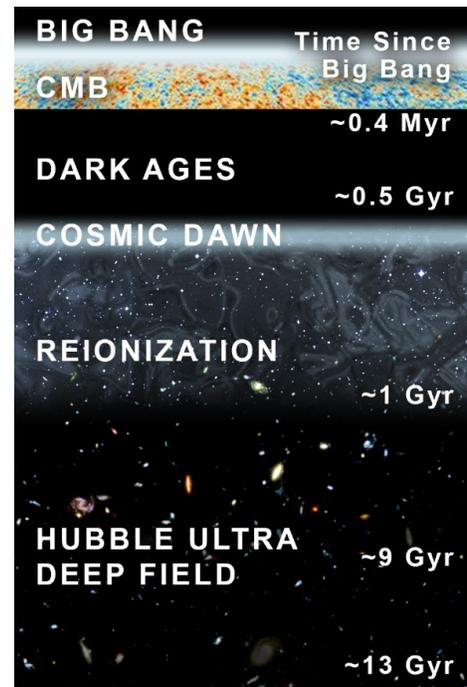

**Figure 1:** The pre-stellar (Dark Ages), first stars (Cosmic Dawn), and Reionization epochs of the Universe can be uniquely probed using the redshifted 21-cm signal. This history is accessible via the neutral hydrogen spin-flip background. Credit: Robin Clarke, Caltech/JPL.

This transformative event marked the first emergence of structural complexity in our Universe, but no currently planned mission can make observations this far back in time. While JWST (*James Webb Space Telescope*), the *Roman Space Telescope*, and a suite of ground-based facilities will observe the Universe as it was ≳ 300 Myrs after the Big Bang (z ≲ 15, and therefore especially focus on the Reionization era when distant galaxies ionized the gas between them, up to about a billion years after the Big Bang), none now contemplate observing the true first stars and black holes [3]—much less the Dark Ages that preceded them. For example, CMB observations of Thomson scattering measure the integrated column density of ionized hydrogen, but only roughly constrain the evolution of the intergalactic medium [4]. Ly$\alpha$ absorption from quasars only confines the end of reionization at relatively late times, z∼7 or 770 Myrs after the Big Bang [5]. Observations with HST (*Hubble Space Telescope*) and JWST will simply find the brightest galaxies at high redshifts (z≲15), and thus any inferences drawn about the high-z galaxy population as a whole depend upon highly uncertain assumptions about the faint-end slope of the luminosity function [6]. The *Hydrogen Epoch of Reionization (EoR) Array* (HERA; [7]) will attempt to observe the neutral hydrogen angular power spectrum over a bandwidth of ∼50–200 MHz (z = 27 − 6, 117 − 942 Myrs) covering the EoR but not the Dark Ages [8], where experiments on Earth have substantial problems (see §1.3).

The void of observations begins at the Dark Ages, with the intergalactic medium (IGM) composed of newly formed neutral hydrogen, extends through Cosmic Dawn starting at z∼30-40, when the first stars, galaxies, and accreting black holes form, and ends in the Universe's last phase transition with an ionized IGM. Models of standard ΛCDM cosmology[1] and primordial star formation are untested during this critical time in the early Universe.

**There is one unique way to probe these early epochs, however, using the redshifted 21-cm background observed from the farside of the Moon.**

### 1.1.2 The Hydrogen Cosmology Signal

The spin-flip, hyperfine transition of neutral hydrogen [9] produces emitted radio emission at 21-cm (1420 MHz) which is redshifted by a factor of $(1+z)$ due to the expansion of the Universe (e.g., $\nu_{observed}$=18 MHz at $z$=78). **Since neutral hydrogen is ubiquitous during the Dark Ages and Cosmic Dawn, its Global 21-cm spectrum can be measured even by a modest telescope on or above the lunar farside.** This telescope is a precursor for a radio interferometric array to measure spatial fluctuations in the Dark Ages when structure formation is still in a linear mode leading to new constraints on neutrino mass and inflation[2].

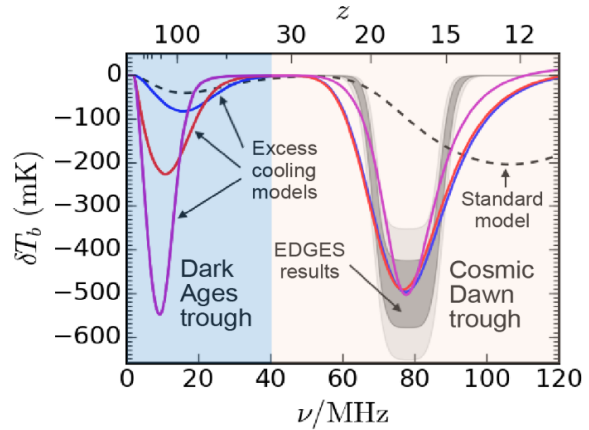

**Figure 2:** The Global 21-cm spectrum provides a key test of the standard ΛCDM cosmology and possible exotic physics produced by interactions with dark matter, as well as constraints on the properties of primordial stars. The black dashed curve is a prediction using standard cosmology with adiabatic hydrogen gas cooling and star formation similar to that in the Milky Way. The color curves are parametric models with added gas cooling [11]. The grey curves are 1- and 2-$\sigma$ uncertainties claimed by EDGES [12] from ground-based observations.

Figure 2 displays the broad spectral features that are common to virtually all Global 21-cm models of the early Universe. Its diagnostic trait, brightness temperature ($\delta T_b \propto$ radio power per frequency bandwidth), is driven by the evolution of the ionization fraction ($x_{HI}$) and spin temperature $T_S$ (measure of the fraction of atoms in the two spin states) of HI relative to the radio background temperature $T_R$ (usually assumed to be the CMB) as given by [1] $\delta T_b \propto \bar{x}_{HI}\left(1 - \frac{T_R}{T_S}\right)$. Higher redshifts correspond to earlier cosmic times. The redshifted 21-cm spectrum is a powerful measure of the time/redshift evolution of structure growth in the early Universe (see Figure 2).

The Dark Ages trough is purely cosmological and thus relatively simple since there are no stars or galaxies during this epoch. The standard cosmology model (black dashed curve) makes a precise prediction of the frequency (≈18 MHz) and $\delta T_b$ (≈40 mK) of the minimum of the trough since only adiabatic expansion cools the gas relative to the CMB. **Any deviation from this prediction is evidence of new physics** (e.g., dark matter annihilation, warm or fuzzy dark matter).

The higher frequency Cosmic Dawn trough is produced by the onset of the first stars. These stars radiate prodigious amounts of Lyα photons that couple with the hyperfine transition of neutral hydrogen and produce an absorption feature via the Wouthuysen-Field effect [1]. X-ray heating of the IGM from the first accreting black holes, probably remnants of the first stars, produces the extremum at the bottom of the trough. **The frequencies of the inflection point near z~30 (see Figure 2)** and the extremum of the trough are measures of the redshifts or times when stars and black holes, respectively, first turn on, and the shape constrains the mass of the first stars and black holes [10]. At present, the parameters of the first radiating objects in the Universe are observationally unknown.

---

[1] The parameterization of the standard cosmological model assumes that the Universe has three components: (1) a cosmological constant (Λ) that is associated with the production of a late time accelerated expansion of the Universe generally attributed to Dark Energy, (2) cold dark matter (CDM), and (3) ordinary (baryonic) matter.
[2] See RFI whitepaper by Burns, Hallinan, Chang et al. entitled "A Lunar Farside Radio Frequency Array for Dark Ages 21-cm Cosmology."

## 1.2 NEW PHYSICS IN THE DARK AGES?

Bowman et al. [12] recently claimed a possible detection of the Cosmic Dawn trough using the EDGES (*Experiment to Detect the Global EoR Signal*) radio telescope in Western Australia. As shown in Figure 2, the EDGES results dramatically differ from the expectations of the standard model. Although the frequency of the bottom of the trough can be understood within the uncertainties of the standard model of star formation, the depth is ≈3 times greater than expected, well below what was thought to be a hard limit set by adiabatic expansion cooling. This result might be theoretically explained by an increase in the radio background above the CMB produced, for example, by synchrotron emission from early star-forming galaxies or quasars [13], or dark matter annihilation [14]. Alternatively, the enhanced trough could be caused by added cooling of baryons via Rutherford-like scattering with partially-charged dark matter [see e.g., 15,16], suggesting several different forms of dark matter. In either case, new and likely exotic physics, is required if this EDGES trough is confirmed.

Figure 2 illustrates parametric added hydrogen cooling models with amplitudes consistent with the EDGES results. The Cosmic Dawn trough originates from a combination of complex physics and astrophysics. On the other hand, the Dark Ages trough is much simpler since this epoch lacks astrophysics. The cooling models nicely separate at these lower frequencies and constrain deviations from standard ΛCDM cosmology that are caused by new physics such as nongravitational interaction with dark matter.

A large number of papers have been published on possible theoretical explanations of the enhanced absorption claimed by EDGES. On the other hand, there are good reasons to be skeptical of the EDGES results based upon possible instrumental systematics [17], beam chromaticity, or other unmodelled effects (see [18,45,46] and references therein, and §1.3). Measurements from the lunar farside provide a timely opportunity to test the reality of the purported trough by conducting observations in the stable, radio-clean environment of the Moon's farside. Even if the EDGES absorption trough is not confirmed, **observations of the redshifted 21-cm line in the Dark Ages and Cosmic Dawn have the potential for discovery of new physics in an unobserved epoch. This includes the decay and/or annihilation of dark matter [19], warm dark matter [29], primordial black holes [20], cosmic strings [21], and early Dark Energy [22].**

## 1.3   THE ADVANTAGES OF THE LUNAR FARSIDE

Because of the high dynamic range necessary to see the hydrogen spectrum against the astrophysical foreground (Figure 3), the main concern of 21-cm Global signal experiments is instrumental and environmental stability. While the instrumental aspects of stability are described in §2 (Table 1), environmental stability is also crucial. **Access to the lunar farside provides an unparalleled opportunity to perform precision cosmology due to the dry, unchanging landscape, the lack of a significant ionosphere, and the unique radio-quiet environment.** The long-term stability of the lunar environment enables extended integration and calibration to gather precise measurements of astrophysical and cosmological parameters.

Conditions on Earth significantly affect the antennas for Global signal experiments. Variations in ground and atmospheric conditions can modify the radiation coming from the sky and the antenna's electrical properties. Changing humidity in the air and soil (which alters the dielectric constant), seasonal changes in vegetation, and variations in temperature (expansion/contraction of the antenna) modify the beam pattern in unknown ways which complicate the extraction of the 21-cm signal against the foreground. By observing from the lunar surface during the day and/or night, once the temperature equilibrates, a Global 21-cm telescope will have superbly stable environmental conditions.

Earth's ionosphere presents another major challenge to terrestrial experiments. The D-layer absorbs radiation and the F-layer refracts it. These effects generate spectral systematics that cannot be accounted for without real-time knowledge of the ionosphere at a level not currently achievable. Even if the ionospheric properties (e.g., electron density) are well-known, a direction-dependent calibration scheme would be necessary to correct for it. As such a scheme has not yet been devised, Earth-based experiments cannot currently perform precision measurements of the Global 21-cm spectrum. This argument is persuasively described in Shen et al. [24; see also 25,26] on behalf of the Cambridge REACH (*Radio Experiment for the Analysis of Cosmic Hydrogen*) team [27]. They conclude, "it is clear that the ionospheric effects have great potential to bury a weak Global 21-cm signal of amplitude ≲0.6 K." By observing from the Moon, a Global 21-cm telescope is free from all ionospheric effects at our observation frequencies.

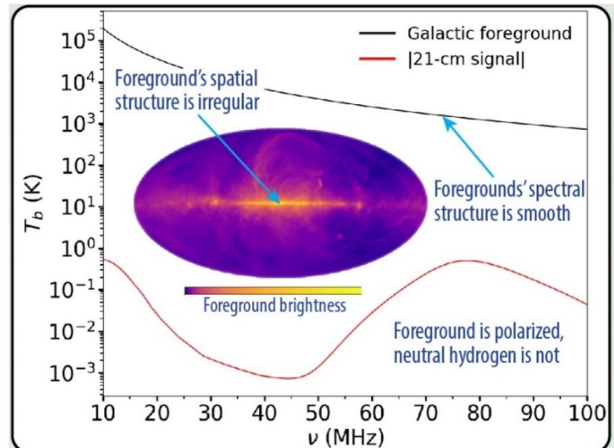

**Figure 3:** The 21-cm global spectrum (red curve) must be measured in the presence of a bright Galactic foreground (black curve). The foreground is separated from the signal by using differences in spectral shapes, spatial structure, and polarization. Note the log-linear plot in contrast to Figure 2. *Insert:* Mollweide projection of the sky at 408 MHz [36].

Any Global signal experiment on Earth or in Earth-orbit experiences anthropogenic Radio Frequency Interference (RFI) caused by powerful military and commercial transmitters. Excising RFI localized in one frequency channel that contains much more power than adjacent channels is simple enough. However, this strategy is insufficient to remove all relevant RFI both because there is low-level leakage across adjacent channels and because some RFI (e.g., over-the-horizon RFI bouncing off the ionosphere) is invisible to such straightforward techniques. Being in view of Earth at the distance of the Moon, such as from the lunar nearside, does not help because the brightness temperature produced by terrestrial RFI is 750,000 K [28].

The lunar farside is the only location within the inner solar system from which to avoid terrestrial RFI, with most of the farside having suppressed RFI in excess of -80 dB level at >10 MHz according to our electrodynamic modeling [28]. Most of the lunar farside consists of highlands with well-mixed subsurface [e.g., 40]. The best conditions for our radio measurements would be subsurface down to a few electrical skin depths (e-folding penetration depth of radio waves) at 60 MHz (3-9 m) with modest substructure no larger than a wavelength (5 m). This would produce only slight scattering of radio waves and should not complicate the antenna beam (see Figure 9). However, we can potentially deal with subsurface structure by measuring scattering using passive lunar sounding, with powerful solar Type II/III radio bursts as sources, following an approach pioneered by Apollo 17 [41] and correct our beam model set accordingly.

## 1.3 SCIENCE OBJECTIVES

The instrument requirement of low frequency radio observations from the lunar farside is to measure the two absorption troughs in the redshifted 21-cm spectrum (Fig. 2). They will resolve current ambiguities and cleanly constrain the origin and characteristics of added baryonic cooling and/or an enhanced radio background. The scientific objectives are:

1. Determine the level of (dis)agreement with the standard cosmological model possibly caused by dark matter or other exotic physics in the Dark Ages.
2. Determine (a) the level of excess cooling above the adiabatic limit for Cosmic Dawn and (b) when the first stars and black holes formed.

These objectives will broadly address the key scientific question: **Will the observed behavior of the HI in the early Universe redefine the standard cosmological model and reveal new physics?**

## 1.4   RELEVANCE TO NASA ASTROPHYSICS & DOE HIGH ENERGY PHYSICS PRIORITIES

A Global 21-cm telescope on or in orbit above the lunar farside addresses NASA's objectives in astrophysics to "Discover how the universe works, explore how it began and evolved" and to "Probe the origin and destiny of our universe, including the nature of black holes, dark energy, dark matter and gravity." It will execute a mission recommended in the NASA Astrophysics Roadmap to observe "the Universe's hydrogen clouds using 21-cm wavelengths via observations from the farside of the Moon." The recent Artemis III Science Definition Team (SDT) Report advocated for observations from "the farside of the Moon and the 21 cm electromagnetic radiation spectral line to study the Dark Ages of the universe."

The NRC Astro2010 Decadal Survey, New Worlds, New Horizons in Astronomy and Astrophysics, identified the Dark Ages and Cosmic Dawn as one of cosmology's great frontiers. A key question posed by the survey, "What were the first objects to light up the Universe, and when did they do it?" will be addressed by the telescope.

Redshifted 21-cm observations of the Dark Ages and Cosmic Dawn are consistent with the mission of DOE HEP "to understand how our universe works at its most fundamental level". These observations support three of the Science Drivers of Particle Physics including "identify the new physics of dark matter", "understand cosmic acceleration: dark energy and inflation", and "explore the unknown: new particles, interactions, and physical principles."

## 2.   Global 21-cm Telescope for the Moon

We recently completed a NASA-funded engineering design study for a first Global 21-cm telescope on the Moon called DAPPER[3] (Dark Ages Polarimeter PathfindER). Here, we describe how DAPPER will measure the 21-cm spectrum using straight-forward technologies in preparation for more advanced experiments, much the same way as COBE's first wide-band measurements of the CMB. Table 1 lists the top-level requirements and physical properties of our telescope.

**Table 1.** Summary of Global 21-cm Telescope Instrument and Mission Properties

| Science Measurement | Instrument Requirements | Mission Requirements | Instrument Mass | Instrument Power | Instrument Volume |
|---|---|---|---|---|---|
| • Brightness Temperature<br>• Redshifts: 12-140 | • Frequencies: 10-110 MHz<br>• Spectral resolution: 50 kHz<br>• Thermal noise: 20 mK<br>• Receiver gain: 5 ppm/sec | • Observations from lunar farside<br>• Integration time ≳1 lunar day<br>• S/C-lander EMI < $10^{-12}$ V/(Hz)$^{1/2}$<br>• Receiver temperature variations: $\pm 1°C$ | 32 kg including 23% contingency | 45 W including thermal management + 30% contingency | • Patch antenna: 0.85×0.85×0.45 m<br>• Electronics package: 0.6×0.6×0.5 m |

### 2.1   TELESCOPE DESIGN APPROACH

The major challenge of hydrogen cosmological observations from the Moon is the presence of an astrophysical foreground that is much brighter than the 21-cm signal. Our strategy to separate the 21-cm signal takes advantage of three distinct characteristics of the foreground in contrast to the signal.

---

[3] https://www.colorado.edu/project/dark-ages-polarimeter-pathfinder/

- As opposed to the broad frequency structure of the 21-cm signal, the frequency spectrum of the foreground is a simple, low-order polynomial (power-law) produced by synchrotron radiation (e.g., [30,31,32]) — see Figure 3.
- The Galaxy has significant spatial structure on the sky whereas the cosmological signal is isotropic on scales ≳10°.
- In addition to intrinsic sky polarization, spatial structure in the foreground induces a polarization response in the dipole antenna, whereas the uniform, unpolarized 21-cm signal produces a response only in Stokes I. Measurement of the four Stokes parameters enables a clean separation of the primordial signal from the foreground [10,33].

This strategy incorporates two features that greatly enhance the probability of a detection. These include (1) dynamic polarimetry, a method for separating the foreground spectrum from that of the HI [34,35] using an approach similar to that employed for CMB observations, and (2) application of a pattern recognition methodology that characterizes signals and systematics via modeling sets so that the 21-cm signature can be identified [33], also similar to that used for the CMB.

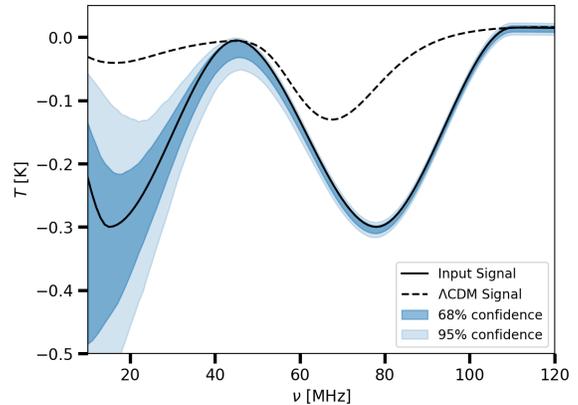

In Figure 4, the precision with which the 21-cm brightness temperature is expected to be measured using the above strategy is shown for a telescope on the lunar surface operating over one lunar day. This end-to-end instrument simulation includes the major sources of noise and systematics, including the sky, the beam-weighted foreground, the lander, and lunar subsurface. Our data analysis pipeline using pattern recognition, statistical inference, and modeling sets was employed to extract the spectrum from the foreground. Further details on the instrument and pipeline used to construct Figure 4 are given in §2.2 and §2.3, respectively.

**Figure 4:** This end-to-end Global 21-cm cosmology telescope simulation (see §2.3), including thermal noise along with instrument and foreground systematics, illustrates a clean >5σ separation between the standard cosmological model (dashed black line) and an added cooling model consistent with EDGES (blue). From the shape, depth, and frequencies of the troughs, the cooling rate and the redshift at which cooling commences ($z_0$) are calculated [23]. 240 hours of integration during the lunar day was assumed; further integration during the lunar night will shrink the uncertainty bands.

### 2.2 TELESCOPE INSTRUMENTATION

The required instrumentation is relatively simple with ground-based and component flight heritage. The wire or STACER (Spiral Tube & Actuator for Controlled Extension Contraction) antennas for the low frequency part of the band have deployed and flown successfully on many space science missions. The Patch antenna for the high end of the band is an engineering extension of those flown extensively in space, mostly for down-link communications. See Figure 5. The four-channel correlation spectropolarimeter is commonly used in radio astronomy and can be embedded on a single electronics board.

#### 2.2.1. Antennas

A single antenna cannot span the wide band of observations from the Dark Ages to the Cosmic Dawn over 10-110 MHz. But this can easily be done with two antennas, both of which are much simpler than that needed, for example, for mm-wave observations of the CMB. An illustration of the two antennas is shown in Figure 5.

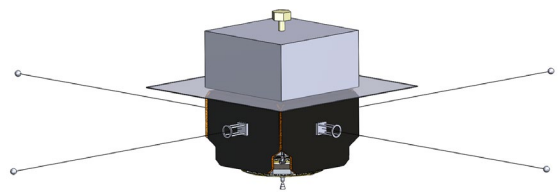

**Figure 5:** Global 21-cm cosmology observations over a wide band from 10-110 MHz can be made using two, straight-forward antennas – thin wire or STACER antennas deployed from the telescope body and a non-deployable Patch antenna (grey cube).

The low band (~10-45 MHz) antenna subsystem consists of 4 deployable wire boom monopoles, arranged in two orthogonal co-linear dipoles. The deployment is expected to be 5-7 m tip-to-tip. The antenna has limited directivity, but this is not required for an all-sky experiment. The antenna subsystem meets the requirement for dual polarization. These boom antennas are similar to those flown on the NASA Wind, Van Allen Probes, and THEMIS spacecraft. This antenna system is at TRL 8.

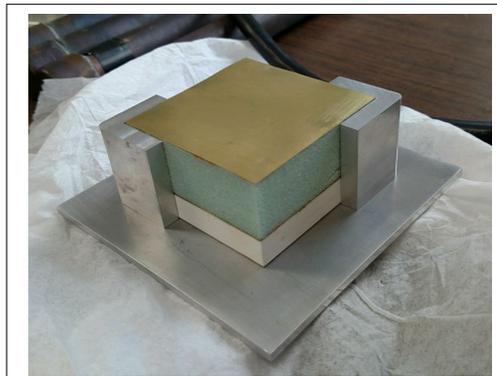

**Figure 6.** This 1/15th-scale Patch antenna is a stacked structure with two square pieces of thin brass sheet stock separated by dielectric materials. It is attached to a thicker aluminum plate that serves as a rigid base. Epoxy is used to bond the layers together. The upper dielectric material is extruded polyurethane. The lower dielectric material is Cumming Microwave C-Stock, a ceramic loaded polymer engineered to produce a specific dielectric constant. Coaxial RF connections are attached under the base, with the center conductor extending through the lower dielectric and bonded to the central metal sheet. The antenna performance is consistent with CST model predictions. *Image: NRAO.*

A dual-linear polarization, rectangular Patch antenna will be used because of its inherently smooth and consistent beam power pattern over the 40-110 MHz high band. It is a microstrip-type antenna and an excellent choice where size, mass, performance, ease of fabrication, and aerodynamic profile are constraints [42]. Conceived in 1953 [43], the microstrip antenna has been studied carefully and used extensively over the years in a variety of applications. Intrinsically a high Q structure, the operational bandwidth has been extended employing a stacked profile [44]. For many years, the Patch antenna has been used for communication and navigation functions onboard spacecrafts. For example, a 36 element, broadband, circularly polarized array of Patch antennas, operating at 1.0-1.5 GHz, was used successfully for navigation onboard GIOVE-A[4]. The DAPPER Patch is essentially a scaling of this element design by a factor of fifteen with engineering modifications to extend the bandwidth slightly and accommodate interface requirements.

For the Patch antenna, we simulated 1/15th and 1/4th scale models with various dielectric designs (solid, graded, composite) in order to optimize performance vs. overall mass. These smaller versions permit full characterization of the design in a cost-effective manner. The first 1/15th scale model was recently fabricated by NRAO (Figure 6) and is currently undergoing radiometric tests at 600-1650 MHz. For the electronics package, NRAO built and tested proof-of-concept boards to confirm the calibration correlations associated with cross-polarization. This development work yields an expected TRL 4 by mid-year and TRL 5 by the end of 2021.

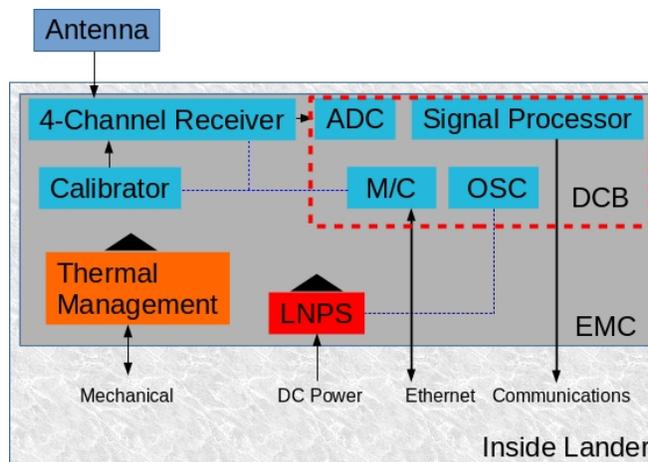

**Figure 7:** The DAPPER high-band instrument shown in this high-level block diagram meets the requirements in Table 1 to produce precision measurements of the Global 21-cm spectrum from 40-110 MHz. These data will be combined with low-band 10-45 MHz measurements using similar electronics.

---

[4] https://earth.esa.int/web/eoportal/satellite-missions/g/giove-a

### 2.2.2 Receiver and Electronics

A top-level block diagram of the telescope is shown in Figure 7 together with the spacecraft/lander interface connections. The tightly integrated instrument package consists of the antenna, four-channel receiver, calibrator, analog/digital converter (ADC), digital signal processor, and monitor/control (M/C) system, master oscillator/clock, and power conditioning. A cross-sectional sketch is given in Figure 8. The Patch antenna protrudes above the top surface of the spacecraft/lander (which forms the electromagnetic reflector for the antenna), and the electronics is housed inside the spacecraft/lander. Inside the package, Compartment #1 contains the receiver and noise calibrator, Compartment #2 the tone calibration, Low Noise Power Supply (LNPS), and the thermal control system. Compartment #3 contains the Data Control Board (DCB). The DCB is the command and data interface between the instrument and the spacecraft/lander. It receives and implements commands, via the interface specified by the lander. This board contains four functional units: master oscillator/clock, Analog-to-digital converters, signal processor, and monitor/control. Instrument stability and signal purity are of critical importance for this instrument and these are controlled via the LNPS, thermal management of the analog and ADC electronics, and adherence to electromagnetic compatibility (EMC) plans throughout.

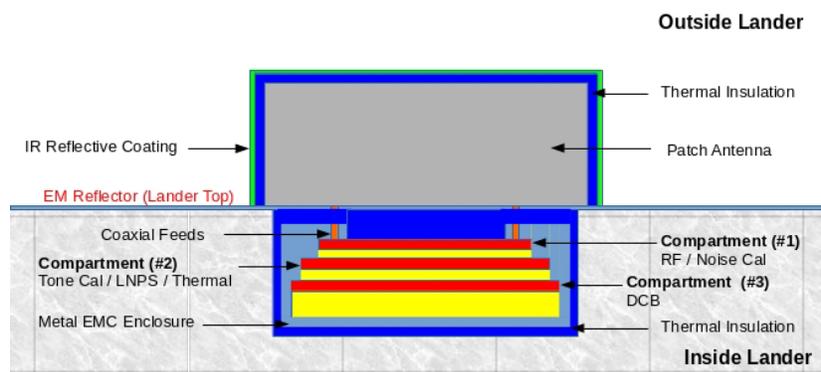

**Figure 8:** Cross-section of the DAPPER instrument shows a compact, integration between the Patch antenna and the electronics package.

### 2.2.3 Performance

The EM performance for the Global 21-cm telescope on the lunar surface was modelled using CST Studio Suite[5]. A Patch antenna was placed on the top deck of a lander at 2.4-m above the lunar surface (using the Astrobotics Peregrine lander as a model). The top deck serves as a ground screen/reflector (2.5×2.0 m) for the Patch antenna. The interactions of the antenna with the lander, the deployed low band STACER antennas, and the lunar subsurface contribute to the antenna response. The simulation assumed 1) an anhydrous regolith (dielectric constants ranging from 2.5-6, and loss tangents=0.02-0.5), 2) a depth >>electrical skin depth, and 3) a well-mixed subsurface layer. The resulting beam pattern and antenna efficiency for the Patch antenna, shown in Figure 9, are smoothly varying functions of frequency, which enable the separation of the 21-cm signal from the foreground. **The instrument properties described above were used in the end-to-end simulation in Figure 4, which meets the science requirements for our radio observations on the lunar farside.**

---

[5] https://www.3ds.com/products-services/simulia/products/cst-studio-suite/

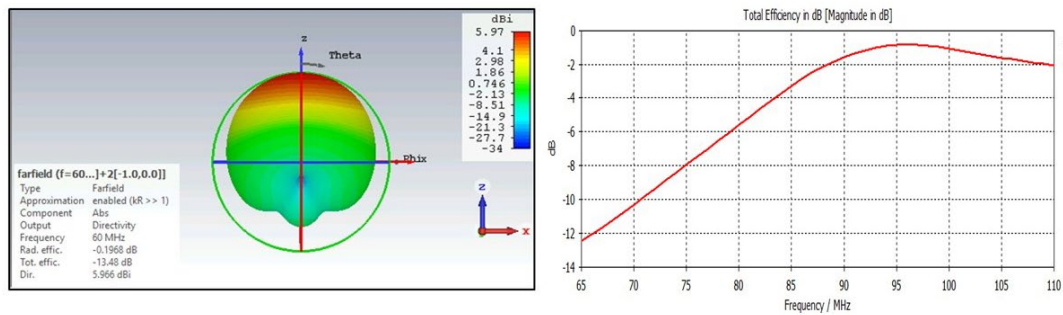

**Figure 9:** CST electromagnetic modeling of the Patch antenna on the top deck of a lunar lander sitting on the farside surface demonstrates good performance characteristics including a smooth beam pattern (example at 60 MHz is displayed with a small backlobe (left) and antenna efficiency that is a continuous function of frequency (right).

### 2.3    THE TELESCOPE DATA ANALYSIS PIPELINE

In order to rigorously separate the small Global 21-cm signal from the large low-radio-frequency foreground, a data analysis pipeline was developed that utilizes well-tested pattern recognition and statistical methodologies to properly account for uncertainties in each of the data components as well as their covariance. The foreground is modeled using independent observations covering the entire sky that are then extrapolated to the data frequency range via spectral index modeling and weighted with suitable antenna beams. The latter are generated using standard electromagnetic simulations in combination with scaled lab measurements. We build-in the associated uncertainties by creating model sets of spectra which encompass them (see e.g., [33,41]). These model sets also include multiple foreground views as a function of observation time or sky region and Stokes parameter. Such modeling allows us to leverage the distinct foreground views from different parts of the sky or polarization to be found in observed spectra against the unchanged isotropic, unpolarized 21-cm signal and robustly extract the signal [37].

Each model set of spectra, for the beam-weighted foreground or the signal, is then decomposed into an optimal vector basis to capture the modes of spectral variation for the given model set, in order of importance. These optimal bases are used to form linear models, where the parameters are the coefficients multiplying each mode, with which to fit the foreground and signal components in the observational data. The numbers of modes employed for each linear model are selected by minimizing an information criterion of choice or the Bayesian evidence (if using priors) when fitting the data with the sum of both models [33]. Importantly, the goodness-of-fit of the model sets can be statistically tested as explained in [38].

In the next step of the pipeline, the constraints on the overall linear model are employed as a starting point to smoothly initiate a numerical Markov Chain Monte Carlo exploration of the full posterior probability distribution of a chosen, non-linear signal model [39]. Results from simulations show meaningful statistical plus systematic uncertainties recovering the input parameters. Following on the success of Bayesian inference as applied to measurements from the CMB, Global 21-cm experiments are incorporating these statistical techniques. Bayesian statistics permit us to self-consistently add prior knowledge and fit systematic and signal parameters in the presence of noise to avoid biases and spurious error estimation. Our pipeline was used to extract the model spectrum in Figure 4 in the presence of realistic foregrounds and instrument systematics.

### 3.    SUMMARY AND RESPONSES to RFI Questions

- **Key Science Topics –** A first, relatively simple, Global 21-cm Cosmology telescope on or in orbit above the lunar farside will open a new frontier to the early Universe, the Dark Ages and Cosmic Dawn, allowing unprecedented tests of the standard ΛCDM model of cosmology and the potential revelation of new physics involving Dark Matter, Dark Energy, and other exotic physics. This matches well with both NASA and DOE science drivers to "understand how the Universe works at a fundamental level".

- **Opportunities and Challenges of Lunar Orbit or Lunar Surface** – NASA-funded concept studies of DAPPER have illuminated the opportunities and challenges in both lunar orbit and on a NASA CLPS lunar lander. Either is a viable first mission.

  A SmallSat with total volume of <1-m$^3$ is capable of delivering a high fidelity 21-cm spectrum in ≈2 yrs of operations with a 30% duty cycle to collect data above the radio-quiet farside[6]. The SmallSat requires its own propulsion system to acquire a lunar orbit of 50×120 km. Challenges include the changing thermal environment during a two-hour orbit (going from day to night) as well as reflected/polarized Galactic and thermal emissions from the Moon.

  The farside lunar surface offers a superbly stable environment with two-week long uninterrupted integration times. One challenge is using low frequency instruments to passively sound the lunar subsurface to correct models for telescope beam distortions caused by subsurface reflections (§1.3). Another is thermal management of the telescope as the ambient temperature ranges from -170℃ to 100℃ on the lunar surface from night to day.

- **Technology Capabilities** – DOE and NASA have each invested in radio telescope technologies and relevant data analysis tools. DOE's Berkeley Lab is the lead center for a next-generation CMB experiment, CMB-S4, partnered with Argonne, Fermi, and SLAC labs. The CMB is most analogous to the 21-cm background, coinciding at the beginning of the Dark Ages, where similar observational techniques, such as dynamic polarimetry, and Bayesian analysis are needed to measure signals against a bright Galactic foreground. Here is where a partnership with DOE could be crucial by drawing on its expertise in calibration, instrument systematics, and statistical inference. Similar to the CMB, Global 21-cm observations require us to disentangle signal and beam-averaged foreground covariances. For its part, NASA, via an Astrophysics SmallSat concept study grant, and SMD Exploration, via a technology maturation cooperative agreement, have invested specifically in DAPPER to advance its TRL level for both low frequency antennas and receivers.

- **Precursor Technology Prototypes and Bottlenecks** – Pathfinder ground-based telescopes, such as EDGES and CTP[7], have played an essential role in advancing antenna and receiver designs needed for a lunar-based instrument. Although ground-based systems are unable to detect the Dark Ages 21-cm signal and are severely limited in measuring the Cosmic Dawn signal (see §1.3), they have pointed the way toward calibration approaches and key data analysis tools (§2.3). Another bottleneck involves measuring and correcting for lunar lander and lunar subsurface conditions which distort the telescope beam. Importantly, the most sensitive observations are during the lunar night. Working together, DOE and NASA could pursue a solution to efficiently (with reasonable mass) power the telescope (e.g., advanced battery technologies, MMRTG), including comms and thermal management, through the lunar night.

  DAPPER is a precursor telescope that will open 21-cm cosmology from the Moon. It needs to fly first to clarify the true spectrum of the Cosmic Dawn and to test the calibration and data analysis strategies for the more difficult Dark Ages. It also paves the way for a powerful radio array on the farside that will measure the 3-D spatial fluctuation spectrum of the Dark Ages and provide exquisite probes of the energy scale of inflation, neutrino species, and forms of dark matter among other exotic physics.

- **Collaborations** – Strong functional teams have been built over the past decade to work on 21-cm Cosmology, bringing together leading space science universities including CU-Boulder, UC-Berkeley, ASU; NASA Centers including ARC, GSFC, and JPL; and the NSF's National Radio Astronomy Observatory. DOE Berkeley Lab is already partnering with UCB's Space Science Lab on projects such as CMB-S4. A DOE/NASA collaboration in this arena could also leverage SLAC's core expertise in RF engineering, including low-noise front-end design, signal calibration, signal processing on custom architectures using FPGAs and ASICs, data acquisition, and on-site low-level data reduction.

---

[6] https://www.colorado.edu/project/dark-ages-polarimeter-pathfinder/mission-concept
[7] https://www.colorado.edu/project/dark-ages-polarimeter-pathfinder/ground-based-pathfinders

The Dark Ages Polarimeter Pathfinder: Cosmology from the Farside of the Moon    DAPPER[18] Tauscher, K., Rapetti, D., Burns, J.O. 2020, Formulating and critically examining the assumptions of global 21-cm signal analyses: How to avoid the false troughs that can appear in single spectrum fits, *The Astrophysical Journal*, 897, 132.

[19] Evoli, C., Mesinger, A., Ferrara, A. 2014 Unveiling the nature of dark matter with high redshift 21 cm line experiments, *Journal of Cosmology and Astroparticle Physics,* 11 (024E) (doi:10.1088/1475-7516/2014/11/024)

[20] Tashiro, H., Sugiyama, N. 2013 The effect of primordial black holes on 21-cm fluctuations, *Monthly Notices of the Royal Astronomical Society,* 435, 3001 (doi:10.1093/mnras/stt1493)

[21] Pagano, M., Brandenberger, R. 2012 The 21 cm signature of a cosmic string loop, *Journal of Cosmology and Astroparticle Physics,* **5** (014) (doi:10.1088/1475-7516/2012/05/014)

[22] Linder, E. V. 2006 Dark energy in the dark ages, *Astroparticle Physics,* **26**, 16.

[23] Mirocha, J., Furlanetto, S. R. 2019, "What does the first highly-redshifted 21-cm detection tell us about early galaxies?", *Monthly Notices of the Royal Astronomical Society*, 483,1980.

[24] Shen, E., Anstey, D., de Lera Acedo, E., Fialkov, A., Handley, W. 2020, "Quantifying Ionospheric Effects on Global 21-cm Observations", *MNRAS,* submitted, arXiv:2011.10517.

[25] Datta, A. et al. 2016, "The Effects of the Ionosphere on Ground-Based Detection of the Global 21-cm Signal from the Cosmic Dawn and Dark Ages, *Astrophysical Journal,* **831**, 6.

[26] Vedantham, H.K., Koopsmans, L.V. 2015, "Scintillation noise in widefield radio interferometry", *MNRAS,* **453,** 925.

[27] de Lera Acedo E., 2019, 2019 International Conference on Electromagnetics in Advanced Applications (ICEAA), pp 0626.

[28] Bassett, N., Rapetti, D., Burns, J.O., Tauscher, K., MacDowall, R. 2020, "Characterizing the radio quiet region behind the lunar farside for low radio frequency experiments", *Advances in Space Research*, **66,** 1265.

[29] Boyarsky, A., Iakubovskyi, D., Ruchayskiy, O., Rudakovskyi, A. and Valkenburg, W., 2019. "21-cm observations and warm dark matter models", Physical Review D, 100(12), p.123005.

[30] Bernardi, G., McQuinn, M., Greenhill, L. 2015, "Foreground Model and Antenna Calibration Errors in the Measurement of the Sky-averaged 21cm Signal at z∼20", *Astrophysical Journal,* **799**, 90.

[31] Sathyanarayana Rao, M. et al. 2017, "Modeling the Radio Foreground for Detection of CMB Spectral Distortions from the Cosmic Dawn and Epoch of Reionization", *Astrophysical Journal,* **840**, 33.

[32] Hibbard, J., Tauscher, K., Rapetti, D., Burns, J.O. 2021, "Modelling the Galactic Foreground and Beam Chromaticity for Global 21-cm Cosmology", *Astrophysical Journal,* in press, arXiv:2011.00549.

[33] Tauscher, K., Rapetti, D., Burns, J.O., Switzer, E. 2018, "Global 21-cm signal extraction from foreground and instrumental effects I: Pattern Recognition framework for separation using training sets", *Astrophysical Journal,* **853**, 187.

[34] Nhan, B., Bradley, R., Burns, J.O. 2017, "A Polarimetric Approach for Constraining the Dynamic Foreground Spectrum for Cosmological Global 21 cm Measurements", *Astrophysical Journal,* **836**, 90.

[35] Nhan, B., Bordenave, D., Bradley, R., Burns, J.O. et al. 2019, "Assessment of the Projection-induced Polarimetry Technique for Constraining the Foreground Spectrum in Global 21 cm Cosmology", *Astrophysical Journal,* **883**, 126.

[36] Haslam, G. Wielebinski, R. Priester, W. 1982, "Radio Maps of the Sky", *Sky and Telescope,* **63**, 230.
12